\documentclass[12pt]{article}

\usepackage{amsmath,amssymb,amstext}
\usepackage{epsfig,graphics}
\usepackage{multirow}
\usepackage{latexsym}
\usepackage{hyperref}
\usepackage{wasysym}

\newlength{\figsize}
\begin{document}

%-------------------

\begin{titlepage}

\vspace*{0.60in}

\begin{center}
{\large\bf Symmetry Breaking In\\Twisted Eguchi-Kawai Models\\}
\vspace*{0.65in}
{Michael Teper and Helvio Vairinhos\\
\vspace*{.25in} Rudolf Peierls Centre for Theoretical Physics,
University of Oxford,\\1 Keble Road, Oxford OX1 3NP, U.K.}
\end{center}

\vspace*{0.50in}

\begin{center}
{\bf Abstract}
\end{center}
We present numerical evidence for the spontaneous breaking of the
$Z_N^4$ symmetry of four-dimensional twisted Eguchi-Kawai models with
SU($N$) gauge group and symmetric twist, for sufficiently large $N$.
We find that for $N\geq 100$ this occurs for a wide range of bare
couplings. Moreover for $N\leq 144$, where we have been able to
perform detailed calculations, there is no window of couplings where
the physically interesting confined and deconfined phases appear in
the reduced model. We provide a possible interpretation for this in
terms of generalised `fluxon' configurations. We discuss the 
implications of our findings for the validity and utility of 
space-time reduced models as $N\to\infty$.

\end{titlepage}

%-------------------

\setcounter{page}{1}
\newpage
\pagestyle{plain}

\section{Introduction}
\label{sect:Introduction}

The conventional way of calculating the properties of the large $N$
limit of SU($N$) gauge theories using lattice methods follows three
main steps \cite{TeperReview2004}: the calculation of some
dimensionless ratios of physical quantities at fixed lattice spacing
$a$ and fixed $N$, for several values of $a$ and $N
(=2,3,4,5,\ldots)$; the continuum extrapolation ($a\rightarrow 0$) of
the resulting lattice values at fixed $N$; the large $N$
extrapolation ($N\rightarrow\infty$) of the resulting continuum
values, assuming a leading correction of order $\frac{1}{N^2}$ (for
pure gauge theories). This way of studying the physics of the
SU($\infty$) theory has been the subject of intense research in the
past decade \cite{TeperReview2004}, and has allowed the accurate
calculation of a number of nonperturbative properties of the pure
gauge theory, such as the string tension \cite{LuciniTeperK}, some
glueball masses \cite{LuciniTeperM}, and the deconfining temperature
\cite{LuciniTeperWengerTc}. However, the numerical simulations
involved become very costly when we consider gauge groups of size
$N\apprge 10$ on reasonably large volumes. So, if one wishes to check
by explicit calculation, at very large $N$, that these extrapolations
are indeed under control, one needs another approach. Fortunately
such an alternative approach exists: it uses an old idea introduced
by Eguchi and Kawai that goes by the name of {\it reduced models}
\cite{EguchiKawai82,DasReview}.

In their original proposal \cite{EguchiKawai82}, Eguchi and Kawai
showed that an exact correspondence can be formally constructed
between the $N=\infty$ limit of SU($N$) gauge theories living on an
infinite $d$-dimensional lattice (the original theory) and the same
theory living on a $1^d$ periodic lattice (its reduced model). The
correspondence states that Wilson loops in the original theory and
its reduced model obey the same Dyson-Schwinger equations, and
therefore it is reasonable to deduce that their numerical values
should coincide (ignoring the possibilities raised by multiple
solutions). However, these loop equations are only identical if we
make two assumptions: 1) the large $N$ factorisation of the physical
observables, and 2) an unbroken $Z_N^d$ symmetry of the reduced
model. While it is reasonable to believe that factorisation is a
universal property of a large class of physical operators in the
large $N$ limit \cite{Factorisation}, nothing prevents the $Z_N^d$
symmetry of the reduced model from being spontaneously broken.
Indeed, the symmetry was found to be spontaneously broken in the
continuum limit of the original Eguchi-Kawai model
\cite{QEK-original}, so the Eguchi-Kawai correspondence is spoiled in
that case. But there are alternative ways of constructing reduced
models of SU($N$) gauge theories that keep the $Z_N^d$ symmetry
unbroken for arbitrarily weak couplings: 1) by quenching the
eigenvalues of the reduced link matrices (quenched Eguchi-Kawai
model, or QEK model) \cite{QEK-original}, or 2) by imposing twisted
boundary conditions on the reduced lattice (twisted Eguchi-Kawai
model, or TEK model) \cite{TEK-original}.

A great amount of work was done in the first half of the 1980s in
trying to extract physical properties of large $N$ theories from
numerical simulations of (mainly) TEK models \cite{DasReview}. A
drawback of these models, however, is that the volume of the lattice
on which the original theory is defined is directly related to the
number of internal degrees of freedom of the reduced model, so very
large gauge groups must be used in order to obtain information that
is sufficiently near the continuum and large volume limits to be
physically relevant. Consequently, due to obvious computational
constraints, the results obtained in the early days lacked accuracy
and, in general, did not allow one to reach reliable conclusions
about the physics of the $N=\infty$ theory (for example, we are not
aware of any  finite-temperature study using this framework that
demonstrated, in a single calculation, a clear separation between the
bulk and deconfining phase transitions). With present day
computational resources, one can easily perform more precise
calculations and for larger gauge groups than before, and this is one
motivation for returning to a careful study of large $N$ physics
using reduced models. There are other motivations for such a study:
first, it allows us to do simulations with much larger $N$ than with
the traditional method, since the number of external degrees of
freedom (volume) is reduced to its minimum; and second, it provides
an alternative method to approach large $N$ physics, with which we
can compare (and hopefully confirm) the numerous results already
obtained via the conventional method. Finally we point to the
importance of reduced models, in particular the twisted version,
in matrix models of noncommutative field theory \cite{NCFT} and
M-theory \cite{MM}.

In this paper, we will summarise the results of a detailed numerical
study of the properties of symmetric-twist SU($N$) TEK models in
four dimensions. In the next section, we provide a short review of 
TEK models. Next, we present the details of our numerical simulations 
and discuss the results obtained from them. The surprise is that we 
find that the $Z_N^4$ symmetry of the TEK model, usually assumed to 
be unbroken for all couplings, is in fact broken spontaneously at 
intermediate couplings when $N$ is sufficiently large. At the same 
time there appears to be a persistent metastability at weak coupling 
that prevents the appearance of a physically interesting phase in
the reduced model. We conclude with a discussion about the
possible origin and the implications of this symmetry breaking.
Some of these results have been presented at SMFT 2006
\cite{SMFT2006}.

%-------------------

\section{TEK model}
\label{sect:TEKModel}

The SU($N$) TEK model in four dimensions is defined by the partition
function on a periodic lattice with a single site,
\begin{equation}\label{def:PartitionFunctionTEK}
Z_\mathrm{TEK}= \int[dU]~e^{-b N S_\mathrm{TEK}}
\end{equation}
where $[dU]\equiv\prod_{\mu=1}^4 dU_\mu$ is a product of SU($N$) Haar
measures, $U_\mu\in$ SU($N$) are the link variables on this reduced
lattice, $b=1/g^2N$ is the inverse bare 't Hooft coupling and
$S_\mathrm{TEK}$ is the action of the TEK model,
\begin{equation}\label{def:ActionTEK}
S_\mathrm{TEK}\equiv S_{\mathrm{TEK}}(n;[U])= \sum_{\mu\neq\nu}^4
\mathrm{Tr}\left( I-z_{\mu\nu}(n) U_\mu U_\nu U_\mu^\dag U_\nu^\dag
\right) \geq 0.
\end{equation}
$S_\mathrm{TEK}$ depends on the reduced link variables $U_\mu$ and on
the twist-tensor, $n_{\mu\nu}= -n_{\nu\mu}\in\mathbb Z_N$, that
labels the twist factor $z_{\mu\nu}(n)=
e^{-i\frac{2\pi}{N}n_{\mu\nu}} \in Z_N$. The twist-tensor defines
uniquely (modulo $SL(4,\mathbb Z)$ transformations) the geometry of
the effective lattice on which the original theory is defined. We
choose to use the standard symmetric twist, $n_{\mu\nu}=\sqrt{N}$,
$\forall~\mu>\nu$, of Gonz\'{a}lez-Arroyo and Okawa
\cite{TEK-original}.
With this twist the TEK model corresponds to an effective symmetric
$L^4$ lattice with  $N=L^2$. Under certain conditions, as discussed
below, the TEK model has the same planar limit as a conventional
SU($N$) lattice theory with a plaquette action, at the same value
of the 't Hooft bare coupling. This restricts the physical
value of $N$ to be the square of an integer.

The lattice operators on the reduced $1^4$ lattice are obtained by
replacing the link variables in their original definition with the
corresponding reduced link variables, and then introducing the
appropriate twists. The $I\times J$ Wilson loop in the
$\{\mu,\nu\}$ plane in the TEK model is then given by:
\begin{equation}\label{def:WilsonLoopTEK}
W_{\mu\nu}(I,J) = z_{\mu\nu}(n)^{IJ} U_\mu^I U_\nu^J U_\mu^{\dag I}
U_\nu^{\dag J}
\end{equation}
and the Polyakov loop in the $\mu$ direction
(on the equivalent $L=\sqrt{N}$ lattice) is given by:
\begin{equation}\label{def:PolyakovLoopTEK}
P_{\mu} = U_\mu^L
\end{equation}

The action of the TEK model in (\ref{def:ActionTEK}) has two distinct
symmetries acting on the link variables: the usual gauge symmetry,
\begin{equation}\label{def:GaugeSymmetryTEK}
U_\mu\mapsto\Omega U_\mu\Omega^\dag, ~~~ \Omega\in\mathrm{SU}(N)
\end{equation}
and an additional $Z_N^4$ symmetry,
\begin{equation}\label{def:ZN4SymmetryTEK}
U_\mu\mapsto z_\mu U_\mu, ~~~ z_\mu\in Z_N
\end{equation}
In order for the TEK model to be able to reproduce the same large $N$
physics as the original theory, a necessary condition is to have
vanishing expectation values for the traces of reduced open lines,
such as $U_\mu^\alpha$ with $\alpha$ not an integer multiple of
$L=\sqrt{N}$. In the original theory this is guaranteed by the fact
that such an open line is not gauge invariant, but in the single-site
reduced model any such operator is trivially gauge-invariant and
could be non-zero. However, unlike Wilson loops
(\ref{def:WilsonLoopTEK}) and Polyakov loops
(\ref{def:PolyakovLoopTEK}), we note that reduced open lines are not
invariant under the $Z_N^4$ symmetry (\ref{def:ZN4SymmetryTEK}). So,
if the $Z_N^4$ symmetry is not spontaneously broken, the expectation
value of open lines will indeed be zero, and the loop equations of
the TEK and conventional theories will be identical in the planar
limit.

In the strong coupling limit, $b\rightarrow 0$, the partition
function in eqn(\ref{def:PartitionFunctionTEK}) is dominated by the
Haar measure $[dU]$; consequently, the expectation value of the trace
of a link variable (and in general of any open line) is automatically
zero, so the model has an unbroken $Z_N^4$ symmetry in that limit. We
call this phase the {\it random phase} and denote it by $Z_N^{4(r)}$.
In the weak coupling limit, $b\rightarrow\infty$, the partition
function in eqn(\ref{def:PartitionFunctionTEK}) is dominated by the
classical vacuum of the TEK model, i.e. by the field configuration
$U_\mu=\Gamma_\mu$ that solves the equation:
\begin{equation}\label{def:TwistEaters}
z_{\mu\nu}(n)\Gamma_\mu\Gamma_\nu\Gamma_\mu^\dag\Gamma_\nu^\dag= I .
\end{equation}
The $\Gamma_\mu$ matrices defined above, also known as {\it
twist-eaters}, saturate the lower bound of the TEK action in
eqn(\ref{def:ActionTEK}). They have the property of being traceless,
as do most of its powers: Tr$(\Gamma^\alpha)=
\delta_{0,(\alpha~\mathrm{mod}~L)}$. In other words, the classical
vacuum of the TEK model has a $Z_N^4$ symmetry that ensures that all
lines (except Polyakov loops) vanish. Small quantum fluctuations
around the twist-eating configuration also keep the $Z_N^4$ symmetry
unbroken. We call this phase the {\it twist-eater phase} and denote
it by $Z_N^{4(t)}$. Extrapolating from the fact that the symmetry is
unbroken in the $b\to 0$ and $b\to \infty$ limits, it is generally
claimed that the $Z_N^4$ symmetry of the TEK model should be unbroken
for all couplings. There is, however, no compelling theoretical
argument justifying such a claim for the intermediate-coupling
regime, although it receives some support from the fact that previous
numerical simulations of the model \cite{DasReview} have not shown
any sign of symmetry breaking.

%-------------------

\section{Results}
\label{sect:Results}

In this section we provide numerical evidence that for sufficiently
large $N$, namely $N\geq 100$, the $Z_N^4$ symmetry of the TEK model
is spontaneously broken. For $N\leq 81$ we see no sign of the
symmetry breaking, which is consistent with the results from
numerical simulations done in the past \cite{DasReview}.

\subsection{Phase structure}
\label{sec:Results:PhaseStructure}

In the large $N$ limit, SU($N$) lattice gauge theories possess a set
of  phase transitions that we would expect to observe in our
simulations of the TEK model, if the Eguchi-Kawai correspondence
between the planar limits of the reduced and parent lattice gauge
theories is in fact valid. First, as we increase $b$ we expect to
encounter a first-order phase transition between the strong and weak
coupling phases of the theory. If the lattice is large enough the
transition is between two confining phases. As we increase $b$
further, the lattice will encounter a deconfining phase transition.
This will occur when the smallest lattice length $L_\mu$ satisfies
$a(b)L_\mu = 1/T_c$, and it is characterised by the Polyakov loop in
that direction acquiring a non-zero vacuum expectation value
\cite{BursaTeper}. As we increase $b$ further there will be a
sequence of three further transitions corresponding to each of the
other three Polyakov loops acquiring a non-zero expectation value.
Finally we are in a phase where the volume is small and the physics
on all length scales is perturbative and there are no further
transitions as $a\to 0$. Although it is not obvious that it should be
so, it has been observed \cite{NarayananNeuberger} that exactly the
same phase structure persists on a symmetric $L^4$ lattice, and hence
should be observed in the TEK model. We now discuss these transitions
in more detail.

The bulk transition is a lattice artifact that occurs, for the
standard plaquette action, at $b=b_B\approx 0.36$ in the large $N$
limit. It separates the unphysical phase at strong lattice coupling
(known as the bulk phase) from the physical phase at weak lattice
coupling, through which one approaches the continuum limit of the
theory. The bulk transition also occurs in the TEK model, in this
case separating the random phase, $Z_N^{4(r)}$, from the physically
relevant twist-eater phase, $Z_N^{4(t)}$; it was accurately
determined in \cite{Campostrini} to have the value $b_B=0.3596(2)$ in
the $N=\infty$ limit.

For large enough lattices, the deconfining transition is located on
the physical side of the bulk transition, separating the low $T$
confined phase (with non-zero string tension, $\sigma\neq 0$) from
the high $T$ deconfined phase (with $\sigma=0$). It is signaled by
the spontaneous breaking of the $Z_N$ symmetry of the Polyakov loop
$P_\mu$ along the direction of the 4-torus with the shortest physical
length, $aL=T^{-1}$, when the temperature $T$ rises above a critical
value $T_c$; it is also accompanied by a small discontinuity in the
average value of the plaquette (reflecting the latent heat of this
first-order transition \cite{LuciniTeperWengerTc}). The critical
coupling $b_c$ at which the deconfining transition occurs scales with
the lattice size $L$, so it is natural to expect that there is a
critical size $L=L_B$ below which the deconfining transition is
overtaken by the bulk transition, and only for $L>L_B$ are the two
transitions decoupled. From the calculations of $T_c$ versus $N$ in
\cite{LuciniTeperWengerTc} we can infer that $L_B \approx 9$. In the
TEK model, we can therefore expect to observe a corresponding
deconfining transition, signaled by the spontaneous breaking of the
$Z_L$ symmetry of the reduced Polyakov loops \cite{LAT2005}, once $N
= L^2 > L^2_B \sim 81$. There is some uncertainty in this value, that
will be enhanced in practice by any metastabilities associated with
these first-order transitions, so it is only at very large $N$ that
we can expect the symmetric-twist TEK model to exhibit a genuine weak
coupling confined phase.

The TEK model is equivalent to a symmetric $L^4$ lattice, with
$L=\sqrt{N}$, and so there is no direction that naturally plays the
role of an inverse temperature. This system has been studied in
\cite{NarayananNeuberger}. One finds that as one increases $b$ from
the confining phase one encounters a sequence of four transitions at
each of which Polyakov loops along one more direction acquire
non-zero expectation values. The first of these transitions occurs
when $a(b)L=T_c^{-1}$ and is indeed just the deconfining transition
(as follows from the observed vanishing of finite volume corrections
when $N\to\infty$ \cite{LuciniTeperWengerTc}). The phases are
labelled in \cite{NarayananNeuberger} by $X$c ($X=0,\ldots,4$), with 
0c the confining phase and 4c the small volume phase. Each of of 
these transitions has a continuum limit and hence a specific physical
interpretation. Therefore, if the Eguchi-Kawai correspondence is
correct, we should also be able to observe these $X$c phases in the
symmetric-twist TEK model, on the physical weak coupling side of the
bulk transition.

\subsection{Simulation details}
\label{sec:Results:Numerical}

We have performed Monte Carlo simulations of the four-dimensional
SU($N$)  TEK model with the conventional symmetric twist tensor
($n_{\mu\nu}=L=\sqrt{N}$, for $\mu>\nu$). The thermalised
configurations at fixed $b$ were generated  using the heatbath
algorithm constructed by Fabricius and Haan
\cite{FabriciusHaan-Heatbath}. As a check, a subset of calculations
were performed with a straightforward (albeit less efficient)
Metropolis algorithm, specifically a modified version of the 
algorithm originally constructed by Okawa for the untwisted model 
\cite{Okawa-MC}. The simulations started either from a classical 
twist-eater configuration, referred to as a cold start, or from a 
completely randomised configuration (in practice one that was 
thermalised at $b=10^{-3}$) referred to as a hot start.

We considered gauge groups $N= L^2= 25,36,\ldots,144$ corresponding
to effective lattice sizes between $L=5$ and $L=12$. For each case,
we performed two different simulations differing in the range of
values of $b$ spanned: in the first run (run A), the interval
$b\in[0.10,2.00]$ was spanned with large steps (varying between
$|\Delta b|=0.01$ and $|\Delta b|=0.02$); in the second run (run B)
we wanted to analyse in more detail the region where the phase
transitions occur, so we spanned the interval $b\in[0.200,0.500]$
with smaller steps, $|\Delta b|=0.005$. The other parameters of the
simulations, like the number of sweeps or the number of measurements,
varied with each case and are summarised in Table
\ref{table:SimulationDetails} for the larger values of $N$.

We calculated the expectation values of several reduced observables,
namely the traces and eigenvalue densities of Wilson loops of several
sizes, of Polyakov loops, of link variables ($U_\mu$) and of some
other open lines ($U_\mu^\alpha$, for $\alpha=2,\ldots,5$).

\subsection{Results for $N\leq 81$}
\label{sec:Results:N<81}

The phase structure of the TEK model for $N\leq 81$ is very simple.
Essentially, there is only one phase transition, at $b\approx 0.36$
\cite{Campostrini}. The transition is strongly first-order, as
demonstrated by its relatively large hysteresis cycle, and in each
one of the two phases the average plaquette fits very well the
strong- and weak-coupling expansions of both the TEK model and
Wilson's lattice gauge theory, as shown in Fig. \ref{fig:SU64_plaq}.
Moreover the location of the transition is consistent with that of
the bulk transition in usual lattice calculations at large $N$. So we
can confidently identify it as being the bulk transition.

\begin{figure}[p]
\begin{center}
\leavevmode
\input{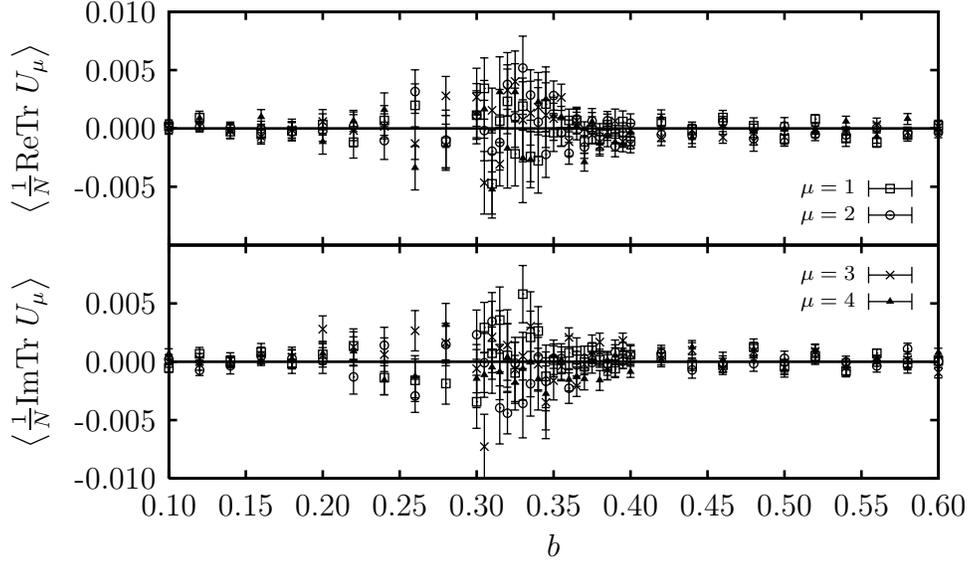}
\end{center}
\vskip 0.015in \caption{Average value of the real and imaginary
parts of traced link variables, $\frac{1}{N}\mathrm{Tr}~U_\mu$, in
the SU(64) TEK model versus the inverse bare 't Hooft coupling, $b$
(from a cold start simulation). \label{fig:SU64_link}}
\end{figure}

We see in Fig. \ref{fig:SU64_link} that the average traces of the
link variables vanish in the whole range of lattice couplings
(although the fluctuations at intermediate couplings are relatively
large). Therefore, we can conclude that the Eguchi-Kawai
correspondence between the planar limits of the reduced model and the
parent (Wilson) lattice gauge theory is valid at least up to $N=81$.
As an example of this, we show in Fig. \ref{fig:SU64_plaq} not only
the 3-loop perturbative result for the plaquette in Wilson lattice 
gauge theory at $N=\infty$, given by \cite{3Loop}: 
\begin{equation}\label{eq:PlaqWeakCoupling-3loop}
\left\langle u_p\right\rangle_W \approx 1-\frac{1}{8b}-
\frac{0.653687}{128b^2}-\frac{0.4066406}{512b^3}+O(b^4)
\end{equation}
but also some values that we obtained from a direct simulation of 
Wilson lattice gauge theory for $4\leq N\leq 16$, extrapolated to 
$N=\infty$. We see from Fig. \ref{fig:SU64_plaq} that these
extrapolated values coincide with the TEK values, within small 
errors that are consistent with the various corrections and 
systematic errors.

The phases separated by the bulk transition can be identified with
the  random phase, $Z_N^{4(r)}$, and the twist-eater phase,
$Z_N^{4(t)}$. The justification comes from an analysis of the
eigenvalue densities of the link variables: in the random phase, the
eigenvalues of the link variables are uniformly distributed over the
unit circle\footnote{Due to unitarity, the eigenvalues of U($N$) or
SU($N$) matrices are constrained to be elements of U(1).}, while in
the twist-eater phase the eigenvalue density of link variables is
composed of $L=\sqrt{N}$ separated lumps spread around the elements
of $Z_L$ on the unit circle (modulo a global $Z_N$ rotation of the
whole spectrum). The latter is due to fluctuations around a
twist-eating configuration (modulo a $Z_N$ rotation), whose
eigenvalues are the elements of $Z_L$ \cite{Billo-DAdda}.

However, for $N\leq 81$, it is not possible to pinpoint a deconfining
transition in the twist-eater phase, since the Polyakov loop and the
plaquette suffer a discontinuity only at the bulk transition. This
result should not be too surprising, since, as remarked earlier, the
sizes of the effective lattices involved, $L\leq 9$, are probably
smaller than the critical size $L_B$ needed to have a clear
separation between the deconfining and bulk transitions. Since
Polyakov loop expectation values are all non-zero in the weak
coupling twist-eater phase, while the strong coupling phase is
confining, the bulk transition can easily be misinterpreted as being
a deconfining transition, particularly in calculations with low
statistical accuracy. The only way to be sure that one has not
confused the two transitions is to vary $b$ so as to identify both
transitions explicitly in a single calculation. As we have remarked
earlier, none of the old TEK calculations that claimed to find a
deconfining transition, displayed such a result. We are therefore
forced to discount those early claims.

This negative result motivated us to extend our investigation to
$L\geq 10$, whose results are summarised next.

\subsection{Results for $N\geq 100$}
\label{sec:Results:N>100}

The phase structure of the TEK model for $N\geq 100$ turns out to be
much richer than for $N\leq 81$, involving several transitions to new
phases. The main characteristic common to all those new phases is
that the $Z_N^4$ symmetry is spontaneously broken, thus violating the
Eguchi-Kawai correspondence.

In the $N=100$ case, the average value of the real part of the
plaquette undergoes two distinct transitions as we decrease $b$ from
a cold-start, but only one transition occurs when we increase $b$
from a hot-start, as shown in Fig. \ref{fig:SU100_plaq}. In the case
of the cold-start scan, the phase in between the two transitions is
characterised by a non-zero expectation value for the trace of one of
the link variables as shown in Fig. \ref{fig:SU100_link_cold} (and
also by an imaginary part of the plaquettes containing that link
variable). This implies that the $Z_N^4$ symmetry of the TEK model is
spontaneously broken to a $Z_N^3$ symmetry in the
intermediate-coupling phase (the traces of the link variables still
have zero expectation value in the other three directions). As
discussed at the end of section \ref{sect:TEKModel}, this effect
should not come as a complete surprise, since there are no compelling
theoretical arguments that prohibit the breaking of the $Z_N^4$
symmetry at intermediate couplings\footnote{That the imaginary
part of the plaquettes can be non-zero follows from the fact that the
action of the TEK model in (\ref{def:ActionTEK}) is not {\it
CP}-invariant. At small $b$ it can be calculated in a strong coupling
expansion and one finds that the contribution decreases with $N$, so
that it is real in the planar limit and is consistent with the
Eguchi-Kawai correspondence.}. The transitions to the regime with
$Z_N^3$ symmetry have a discontinuous character, indicating that
these are first-order phase transitions (or would be if the number of
degrees of freedom were infinite). Increasing $b$ from a hot-start
only one transition is observed, and that is the one from the random
phase to the $Z_N^3$-symmetric phase. As we see in Fig.
\ref{fig:SU100_link_hot}, there is no transition from the $Z_N^3$
phase to the expected weak coupling twist-eater phase, even at very
large values of the inverse lattice coupling. This implies the
existence of stable extrema of the TEK action that are $Z_N^4$
non-preserving and from which we are unable to tunnel because of
large barriers. This immediately raises the question whether this
effect persists for larger $N$ and what consequences it has for the
Eguchi-Kawai correspondence.

As at lower $N$, there is no sign of a deconfining transition in the
physically  relevant phase of the SU(100) TEK model. By `physically
relevant' we mean the twist-eater phase, since it is the only phase
that we observe having  the necessary  $Z_N^4$ symmetry while being
continuously connected to the continuum limit.

\begin{figure}[p]
\begin{center}
\leavevmode
\input{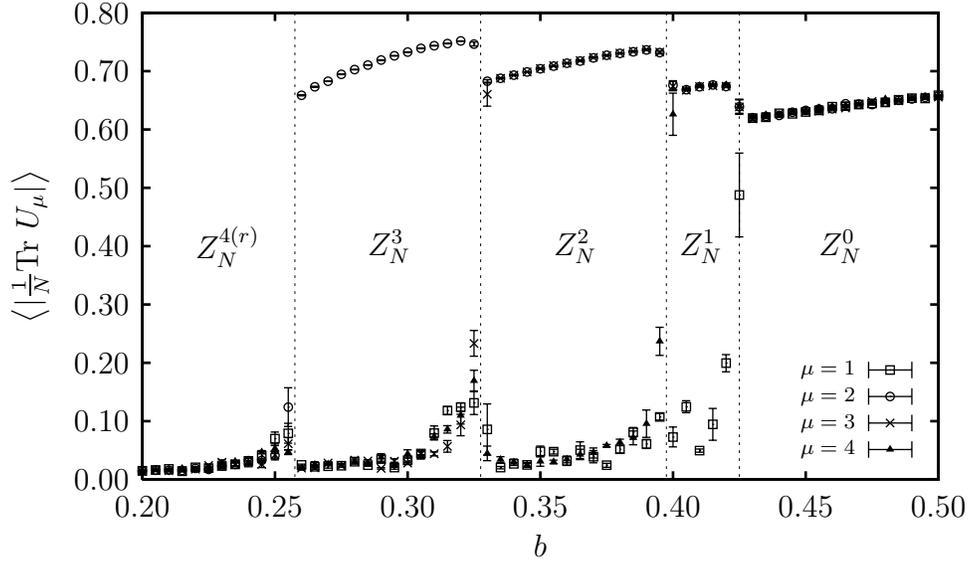}
\end{center}
\vskip 0.015in \caption{Average value of the magnitude of traced
link variables, $\left\langle |\frac{1}{N}\mathrm{Tr}~U_\mu|
\right\rangle$, in the SU(144) TEK model versus the inverse bare 't
Hooft coupling, $b$ (from a hot start simulation).
\label{fig:SU144_link}}
\end{figure}

For $N=121$ and $N=144$ we find, at intermediate lattice couplings, a
sequence of transitions, as illustrated in Fig. \ref{fig:SU144_plaq}
and in more detail in Fig. \ref{fig:SU144_plaq2}. These transitions
are similar to the one we saw for $N=100$, in that each is associated
with the breaking or restoration of one or more of the $Z_N$
symmetries. Thus they lead to non-zero expectation values for the
corresponding link variables, as shown in Fig.\ref{fig:SU144_link}.
We can label the phases by the number $k$ of preserved $Z_N$
symmetries, as $Z_N^k$ ($k=0,\ldots,3$). In the $N=144$ case, all
transitions between different $Z_N^k$ phases are clearly separated,
while for $N=121$ the simultaneous breaking/restoration of several
$Z_N$ symmetries can occur. Tables \ref{table:TransitionsRand} and
\ref{table:TransitionsCold} summarise the critical couplings at which
the observed transitions occur.

\begin{figure}[p]
\begin{center}
\leavevmode
\input{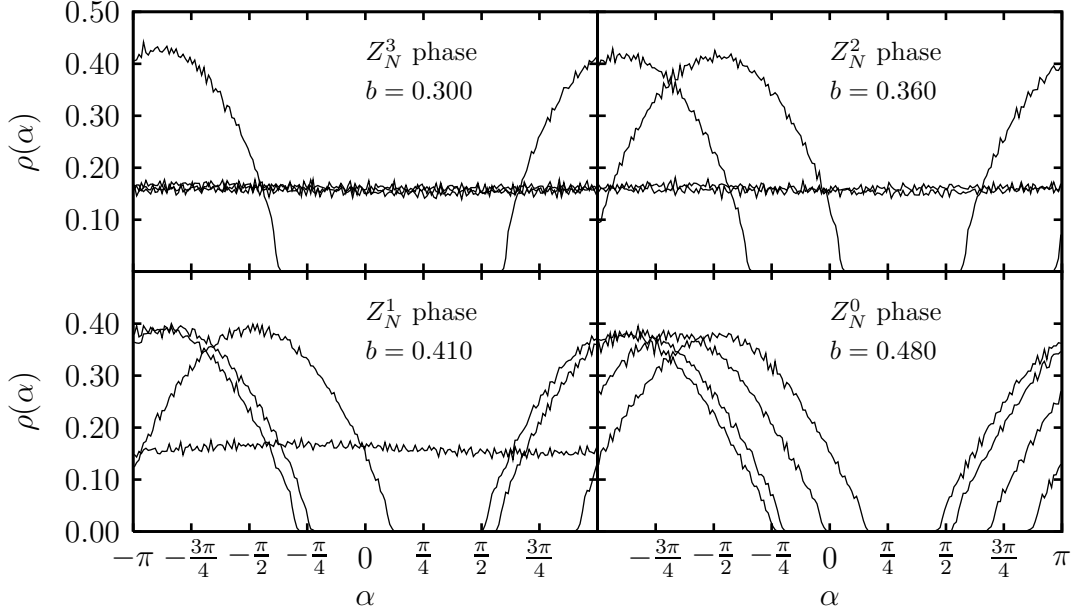}
\end{center}
\vskip 0.015in \caption{Typical profile of the eigenvalue
densities of the link variables, $U_\mu$, in the several
$Z_N^4$-breaking phases of the SU(144) TEK model; $\alpha$ labels the
eigenvalues of SU($N$) matrices, which are pure phases of the form
$e^{i\alpha}$. \label{fig:SU144_link_eig}}
\end{figure}

\begin{figure}[p]
\begin{center}
\leavevmode
\input{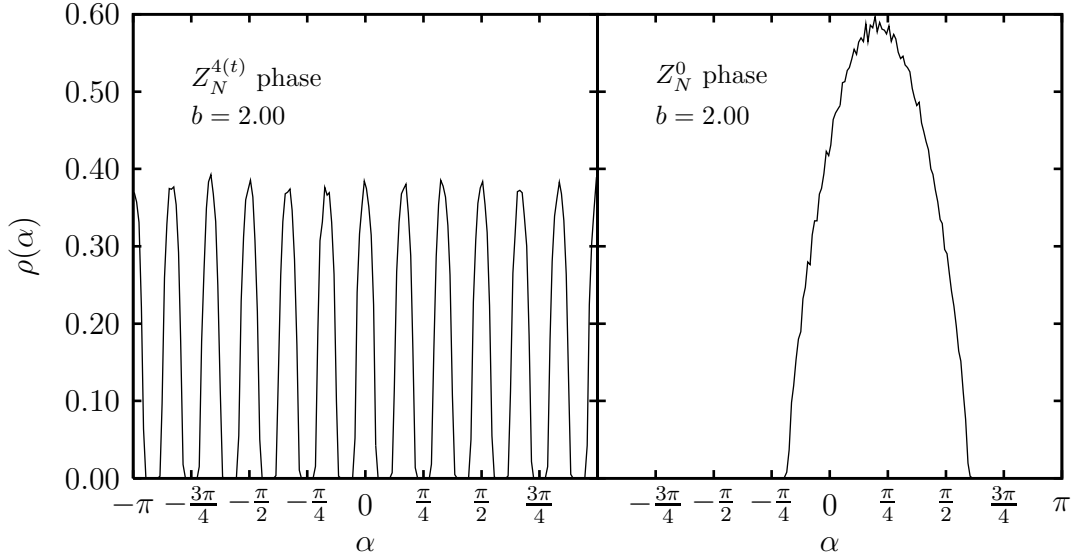}
\end{center}
\vskip 0.015in \caption{Eigenvalue densities of a particular
link variable at an inverse bare 't Hooft coupling of $b=2.00$,
sitting at two different phases of the SU(144) TEK model, namely the
twist-eater phase ($Z_N^{4(t)}$, in the left) and the completely
asymmetric phase ($Z_N^0$, in the right).
\label{fig:SU144_link_eig_compare}}
\end{figure}

Just as for $N=100$, when we increase $b$ from strong coupling for
the SU(121) and SU(144) TEK models, we observe no tunnelling from the
final $Z_N^0$ phase to a twist-eater phase. This reinforces the idea
that some stable extrema of the TEK action are contributing to the
breaking of the $Z_N^4$ symmetry. The nature of these extrema can be
inferred from the eigenvalue spectrum of the link variables. In a
$Z_N^k$ phase, the $k$ link variables of the $Z_N$-preserved
directions have a uniform eigenvalue density, while in the other
$N-k$ broken directions their eigenvalue density consists of a
bell-shaped lump with finite support, spread around an element of
$Z_N$, as illustrated in Fig. \ref{fig:SU144_link_eig}. At larger $b$
the lump becomes narrower, as shown in Fig.
\ref{fig:SU144_link_eig_compare}, indicating that in the limit
$b\rightarrow\infty$ all eigenvalues collapse to one value which is
an element of $Z_N$. In sum, the numerical simulations suggest that
the breaking of the $Z_N^4$ symmetry in the TEK model for $N\geq 100$
might be caused by fluctuations around (stable) configurations of the
form $U_\mu \in Z_N$.

%-------------------

\section{Interpretation}

\begin{figure}[p]
\begin{center}
\leavevmode
\input{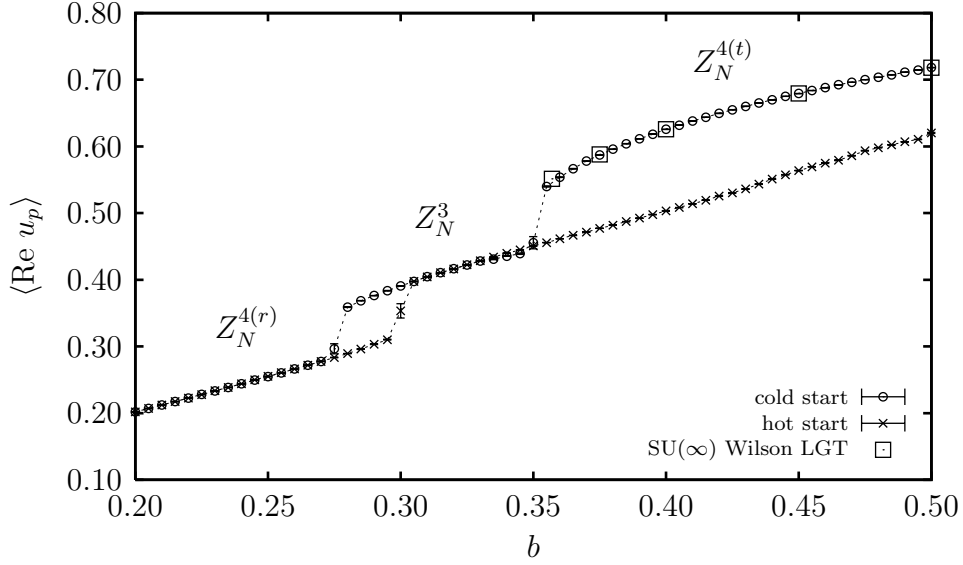}
\end{center}
\vskip 0.015in \caption{Average value of the real part of the
plaquette, $\left\langle\mathrm{Re}~u_p\right\rangle$ in the SU(100)
TEK model versus the inverse bare 't Hooft coupling, $b$. The squares
($\square$) represent the $N\rightarrow\infty$ extrapolation of the
average plaquette in Wilson's lattice gauge theory. 
\label{fig:SU100_plaq}}
\end{figure}

Our numerical calculations suggest that the gauge fields responsible
for the spontaneous breaking of the $Z_N^4$ symmetry are of the form
of fluctuations around $U_\mu\in Z_N$. These field configurations
appear to be able to survive reasonably large fluctuations at
intermediate couplings, and to become increasingly stable as we
increase $b$ to arbitrarily large values. This requires the existence
of stable extrema of the TEK action that break the $Z_N^4$ symmetry.
At the same time we see explicitly from Figs. \ref{fig:SU100_plaq}
and \ref{fig:SU144_plaq} that their action is substantially larger 
than that of the `twist-eating' fields obtained at the same $b$ from 
a cold start. Thus the (infinite?) metastability requires large 
barriers, which implies the existence of corresponding unstable 
extrema of $S_{TEK}$. We will now argue that the elements of $Z_N$ 
may provide such configurations. Our arguments below will follow 
very closely the calculations of  van Baal in his study of surviving 
extrema of the TEK model \cite{vanBaalSurviving}.

Consider a small fluctuation around a general configuration
$\Omega_\mu$, parameterised by an (anti-hermitian) element $X_\mu$ of
the algebra of SU($N$):
\begin{equation}\label{eq:Fluctuation} U_\mu=\Omega_\mu e^{-X_\mu}
\end{equation}
The expansion of the TEK action (\ref{def:ActionTEK}) around
$\Omega_\mu$ up to second-order in $X_\mu$ is given by:
\begin{eqnarray}\label{eq:ActionTEK-2ndOrder}
S_\mathrm{TEK}\approx \sum_{\mu\neq\nu}^4 \mathrm{Tr} \left\{I -
P_{\{\mu\nu\}} - P_{[\mu\nu]}F_{\mu\nu} - \frac{1}{2}P_{[\mu\nu]}
\left( [D_\mu X_\nu + X_\mu,D_\nu X_\mu+ X_\nu] -
[X_\mu,X_\nu]\right) - \right. \nonumber \\ \left. - \frac{1}{2}
P_{\{\mu\nu\}} F_{\mu\nu}^2 \right\} + O(X^3)
\end{eqnarray}
where $P_{\mu\nu}= z_{\mu\nu}(n)\Omega_\mu^\dag \Omega_\nu^\dag
\Omega_\mu\Omega_\nu$ is the reduced plaquette, $P_{\{\mu\nu\}}=
\frac{1}{2}\left(P_{\mu\nu}+P_{\nu\mu}\right)$ is its symmetrisation,
$P_{[\mu\nu]}= \frac{1}{2}\left(P_{\mu\nu}-P_{\nu\mu}\right)$ is its
antisymmetrisation, $D_\mu X_\nu = \Omega_\mu^\dag X_\nu \Omega_\mu -
X_\nu$ is a discretised covariant derivative, and $F_{\mu\nu}= D_\mu
X_\nu - D_\nu X_\mu$ is a discretised field strength tensor. The
stationarity condition for the configuration $\Omega_\mu$ to be an
extremum, $\delta S_\mathrm{TEK} = 0$, then implies the equation:
\begin{equation}\label{eq:Stationarity}
\delta S_\mathrm{TEK}= -\sum_{\mu\neq\nu}^4 \mathrm{Tr}
\left(P_{[\mu\nu]} F_{\mu\nu} \right)= 0 ~~\Longrightarrow~~
\sum_{\mu\neq\nu}^4 \left(\Omega_\nu P_{[\mu\nu]}\Omega_\nu^\dag -
P_{[\mu\nu]}\right)= 0
\end{equation}
An obvious class of solutions to the stationarity condition above are
the configurations satisfying $P_{\mu\nu}\in Z_N$. In fact, van Baal
gave strong arguments \cite{vanBaalSurviving} for the conjecture that
all stable extrema of the TEK action have this form. So let
$P_{\mu\nu}= z_{\mu\nu}(n-m)$, where $n\equiv n_{\mu\nu}$ is the
twist-tensor and $m\equiv m_{\mu\nu}$ is another integer-valued
matrix. Expanding the TEK action around a configuration from this
class reduces (\ref{eq:ActionTEK-2ndOrder}) to:
\begin{equation}\label{eq:Stability}
S_\mathrm{TEK}\approx 2N \sum_{\mu\neq\nu}^4
\sin^2\left[\frac{\pi}{N}(n_{\mu\nu}-m_{\mu\nu})\right] + \frac{1}{2}
\sum_{\mu\neq\nu}^4 \cos\left[ \frac{2\pi}{N}(n_{\mu\nu}-m_{\mu\nu})
\right] \mathrm{Tr} (-F_{\mu\nu}^2) + O(X^3)
\end{equation}
The stability condition, $\delta^2 S_\mathrm{TEK}\geq 0$, is
equivalent to:
\begin{equation}\label{eq:StabilityCondition}
\cos\left[\frac{2\pi}{N}(n_{\mu\nu}-m_{\mu\nu})\right]\geq 0, ~
\forall~\mu,\nu
\end{equation}
since $\mathrm{Tr}(-F_{\mu\nu}^2)= O(X^2)$ is always non-negative.

\begin{figure}[p]
\begin{center}
\leavevmode
\input{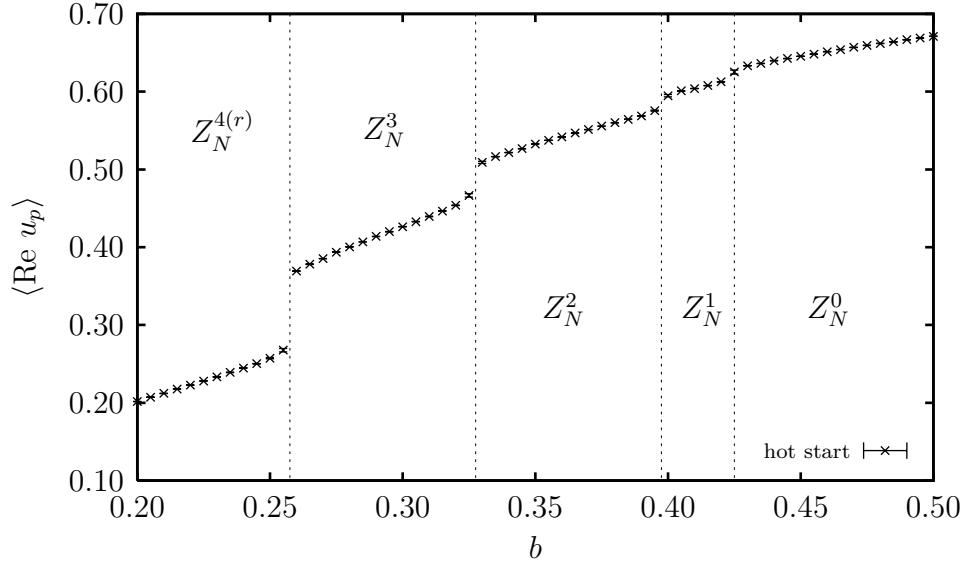}
\end{center}
\vskip 0.015in \caption{Average value of the real part of the
plaquette, $\left\langle \mathrm{Re}~u_p\right\rangle$, in the
SU(144) TEK model versus the inverse bare 't Hooft coupling, $b$
(from a hot start simulation). \label{fig:SU144_plaq2}}
\end{figure}

Let us consider particular solutions to the stationarity condition
$P_{\mu\nu}= z_{\mu\nu}(n-m)$, i.e. $z_{\mu\nu}(m)\Omega_\mu^\dag
\Omega_\nu^\dag \Omega_\mu\Omega_\nu=I$. For the case $m_{\mu\nu}=
n_{\mu\nu}$, the unique solution to $P_{\mu\nu}=I$ is obviously the
twist-eating configuration, the classical vacuum of the TEK model.
Now consider the case when $m_{\mu\nu}\sim O(\sqrt{N})$ and
(\ref{eq:StabilityCondition}) is satisfied. In this case the
solutions to $P_{\mu\nu}=z_{\mu\nu}(n-m)$ (if they exist) are called
{\it fluxons} \cite{vanBaalSurviving}. Fluxons survive the large $N$
limit, being stable minima of $S_{TEK}$, and share some properties
with twist-eaters, like tracelessness; therefore, fluxons are not
expected to contribute to the breaking of the $Z_N^4$ symmetry. Next
consider the case $m_{\mu\nu}= 0$. The solution to
$P_{\mu\nu}=z_{\mu\nu}(n)$ is the set of all SU($N$) diagonal
matrices. In general, these matrices have non-zero trace, and so they
are good candidates for being the stable extrema that break the
$Z_N^4$ symmetry. As a consequence of the first term on the l.h.s. of
(\ref{eq:Stability}), the action of the diagonal matrices grows
proportionally with $N$, i.e. in the partition function such diagonal
matrices appear to be suppressed relative to twist eaters by a factor
of $e^{-bN}$. While naively this appears to be a large suppression,
in a system with $O(N^2)$ degrees of freedom this is not so. Quantum
fluctuations will naturally contribute terms of $O(N^2)$ to the
effective action which can easily overwhelm the difference in the
classical action. At fixed $N$, for large enough $b$, the action
difference eventually dominates; but if the barrier between these
diagonal fields and the twist-eating fields has a natural $O(N^2)$
value, then the tunnelling to the twist-eater phase would be
suppressed by a factor like $e^{-bN^2}$ and would essentially never
occur in a finite simulation -- just as we appear to see in Figs.
\ref{fig:SU100_plaq}, \ref{fig:SU144_plaq} and \ref{fig:SU144_plaq2}. 
Such a barrier will be related to a maximum or saddle-point of the 
TEK action whose action grows $\propto N^2$. Possible candidates for 
such maxima would be configurations satisfying $P_{\mu\nu}=
z_{\mu\nu}(n-m)$, with $m_{\mu\nu}\sim O(N)$ in such a way that the
condition (\ref{eq:StabilityCondition}) is violated.

At the classical level, the set of all diagonal SU($N$) matrices form
a degenerate set of extrema of the TEK action. However, if we
consider the effect of quantum fluctuations, we can easily see that
the degeneracy is lifted and only the elements of $Z_N$ become the
true extrema. This is so because, for arbitrary $X_\mu$, $D_\mu
X_\mu= 0$ for elements of $Z_N$ while $D_\mu X_\mu \neq 0$ for other
diagonal matrices. Consequently the $O(X^2)$ term in
(\ref{eq:Stability}), which is $\propto \mathrm{Tr}(-F_{\mu\nu}^2)$,
vanishes in the former case and is always positive in the latter.
Therefore, when quantum fluctuations $X_\mu$ are taken into account,
the action of a general diagonal matrix becomes larger than the
action of an element of $Z_N$, if $b$ is large enough for $O(X^3) \ll
O(X^2)$, and the degeneracy is lifted. Therefore, the elements of
$Z_N$ provide stable vacua that break the $Z_N^4$ symmetry. This
mechanism is in fact the same as that which leads to the $Z_N^4$
symmetry breaking in the original untwisted Eguchi-Kawai model
\cite{KazakovMigdal}; it is also similar to mechanisms that generate
`order from disorder' in condensed matter physics.

Let us therefore compare the contributions to the TEK action of the
fluctuations $X_\mu$ around twist-eaters ($\Gamma_\mu$) and around
elements of $Z_N$ ($z_\mu$),
\begin{eqnarray}\label{eq:TwistEatersVSCentreFluctuations}
S_\mathrm{TEK}(n;[\Gamma])&=& \sum_{\mu>\nu}^4 \mathrm{Tr}
(-F_{\mu\nu}^2)  + O(X^3)
=  O(X^2) + O(X^3) \\ S_\mathrm{TEK}(n;[z]) &=&
24N\sin^2\left(\frac{\pi}{\sqrt{N}}\right)+ O(X^{3})
=  O(X^0) + O(X^{3})
\end{eqnarray}
We might speculate that the different sensitivities with respect to
the fluctuations $X_\mu$ might be at the origin of the spontaneous
breaking of the $Z_N^4$ symmetry at intermediate couplings. In this
regime the fluctuations are reasonably large, but the Haar measure
does not dominate yet. The contribution to the action of the
fluctuations around twist-eaters is larger ($\sim O(X^2)$) than for
elements of $Z_N$ ($\sim O(X^3)$), so it might be possible that at a
certain critical value of the coupling $b$ the fluctuations around
elements of $Z_N$ will cost less action than the fluctuations around
twist-eaters, and consequently the elements of $Z_N$ become preferred
over the twist-eaters. This argument is, however, very speculative,
since it uses perturbation theory to justify effects that are
observed at intermediate-to-strong couplings, where phase transitions
are also involved.

\begin{figure}[p]
\begin{center}
\leavevmode
\input{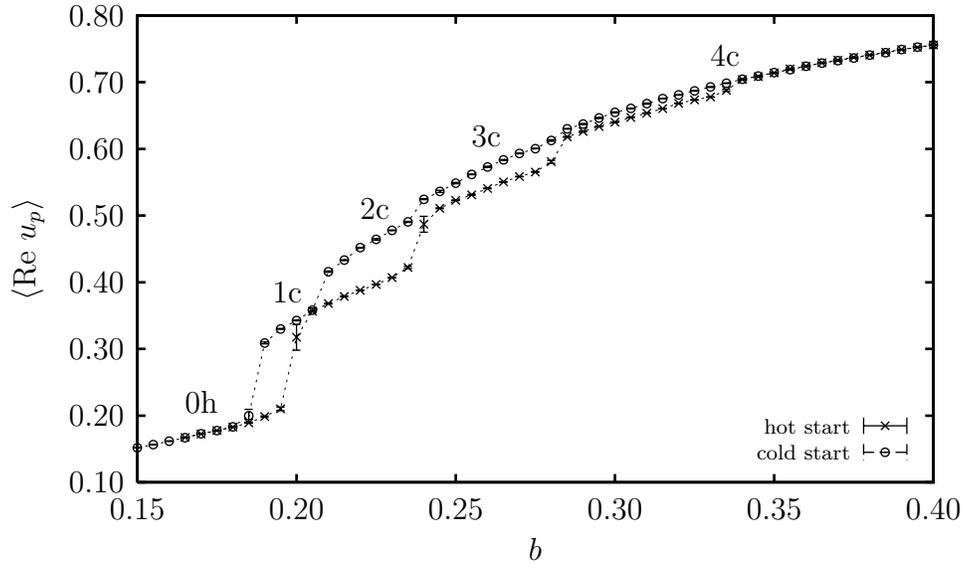}
\end{center}
\vskip 0.015in \caption{Average value of the real part of the
plaquette, $\left\langle\mathrm{Re}~u_p\right\rangle$, in the 
original untwisted SU(81) Eguchi-Kawai model versus the inverse bare 
't Hooft coupling, $b$. The nomenclature used for the different
phases is the same as in \cite{NarayananNeuberger}.
\label{fig:SU81_nEK}}
\end{figure}

Returning to the sequence of transitions observed in the large $N$
TEK model, as shown in Fig. \ref{fig:SU144_plaq2} for $N=144$, we
note their qualitative resemblance to the transitions described in
\cite{NarayananNeuberger} that occur on conventional symmetric
lattices for large $N$ as $b\to\infty$. This might suggest a natural
identification between the $X$c phases and our $Z_N^k$ phases, namely
$X$c $\equiv Z_N^{4-X}$. However, there are at least two
fundamental differences that undermine such an identification. First,
the reduced model version of the so-called 0c phase (the confined
phase on the physical, weak coupling side of the bulk transition,
which is situated between the strong coupling random phase and the 1c
phase where Polyakov loops in one direction have a non-zero
expectation value) was never observed in any of the simulations.
Second, and more importantly, our transitions do not seem to scale
with $L=\sqrt{N}$ in the same way as do the ones in
\cite{NarayananNeuberger}. That is to say, the critical couplings of
the latter increase with the lattice size $L$ in such a way that the
$X$c phases have a proper continuum limit. However, according to
Tables \ref{table:TransitionsRand} and \ref{table:TransitionsCold},
our transitions do not show any sign of the scaling with  $N$ that
would be expected if we were seeing a physical transition on the
equivalent $L=\sqrt{N}$ lattice that occurred at a fixed value of the
physical length  $l = aL = a\sqrt{N}$. That is to say, they appear to
be lattice transitions rather than physical transitions. Moreover all
our $Z_N^k$ phases are situated below the value of $b\approx 0.36$
expected for the bulk transition, and not above (as in the case of
the $X$c phases), and finally their breaking of the centre symmetry
undermines any connection with a supposedly equivalent conventional
lattice. On the other hand, if we forget the Eguchi-Kawai
correspondence (which, after all, we have shown is invalid in this
range of $b$) and consider the TEK model simply as a $1^4$ lattice,
then the links are Polyakov loops, and the identification with the
$X$c phases of \cite{NarayananNeuberger} is essentially trivial. This
suggests looking at the untwisted $1^4$ lattice, where such phases
are known to occur \cite{NarayananNeuberger}. In
Fig.\ref{fig:SU81_nEK} we show a plot of this phase structure
that we have calculated for $N=81$. The phase structure we see
is essentially identical to that obtained in the large $N$ TEK model
with a hot start, as shown in Fig. \ref{fig:SU144_plaq2}, and
the interpretation of the phases appears to be identical. It appears
that for large enough $N$, once quantum fluctuations are sufficiently
important, the effect of the twist factor in the symmetric TEK
model, $z_{\mu\nu}(n)=e^{2\pi i/\sqrt{N}}$, vanishes for all practical
purposes.

%-------------------

\section{Conclusions}

We have analysed numerically the properties of four-dimensional
SU($N$) TEK models with a conventional symmetric twist
\cite{TEK-original}. The gauge groups used in our simulations ranged
from $N=25$ to $N=144$.

For $N\leq 81$, the properties of the TEK model are essentially the
same as those obtained in earlier simulations. As we vary the inverse
bare 't Hooft coupling, $b = 1/g^2N$, we encounter a single
first-order phase transition separating the strong and weak coupling
regions and, crucially, the centre symmetry which is needed for the
Eguchi-Kawai correspondence to be valid, is indeed maintained at all
$b$. However our greater accuracy also makes it clear that there is
no confining phase on the physical weak coupling side of the bulk
transition, in contrast to some claims in the early literature. We
pointed out that this should not occasion great surprise since we
know, from conventional calculations of the deconfining temperature
$T_c$ in SU($N$) gauge theories \cite{LuciniTeperWengerTc}, that the
length of the equivalent lattice, $l=aL=a\sqrt{N}$, is probably less
than $1/T_c$ for all $b$ in the weak coupling region.

For $N\geq 100$, on the other hand, we observe a clear breaking of
the $Z_N^4$ symmetry over a wide range of intermediate couplings,
challenging the conventional wisdom that this symmetry is unbroken
for all couplings. We observe that the breaking of the $Z_N^4$
symmetry occurs step by step in a sequence of transitions, with
intermediate phases in which one or more of the $Z_N$ symmetries are
unbroken. When, for our largest values of $N$, we increase $b$ from
$b\sim 0$ the final weak coupling phase appears to be one in which
all four centre symmetries are spontaneously broken. At large enough
$b$ it becomes clear that the symmetry breaking fields are nothing
but fluctuations around centre elements. These are stable minima of 
the twisted lattice action, as originally discussed by van Baal
\cite{vanBaalSurviving}. Since the action of these `generalised 
fluxons' is only $O(N)$ higher than that of the absolute minimum 
(the twist eater), it is not surprising that it does not tunnel to 
the latter since the generic barrier height will be $O(N^2)$ in an 
SU($N$) gauge theory.

The sequence of symmetry breaking transitions in the TEK model at
very large $N$ is essentially identical to that which occurs in the
original untwisted Eguchi-Kawai model (apart from a shift in the
location of the transitions). It is as if the effect of the twist, a
phase factor $e^{2\pi i/\sqrt{N}}$ associated with each plaquette,
becomes irrelevant for large enough $N$. Naively one might expect
this to happen once the phase is small, i.e. for $2\pi/\sqrt{N} \ll
1$ or, equivalently, $N\gg 4\pi^2$. This provides a very large scale
for what what we mean by $N$ being `large', and might explain why new
effects arise for $N > 100$ in the TEK model, in contrast to
conventional calculations where finite $N$ corrections are,
typically, already very small for $N=6$ or $N=8$.

One might still hope to encounter physically interesting phases by
beginning at $b=\infty$ with a twist-eating field configuration and
then decreasing $b$. If the first transition to strong coupling were
to be at $b\simeq 0.360$, as it is for $N\leq 81$, then the $12^4$
lattice that is equivalent to the SU(144) TEK model, should be large
enough to be in the physical confining phase before reaching that
value of $b$. Frustratingly, however, for SU(144) the critical value
of $b$ increases, as we see in Tables \ref{table:TransitionsRand} and
\ref{table:TransitionsCold}, so that there is once again no window
for any weak coupling confining physics. Whether this will continue
if $N$ is increased further is obviously a very interesting question.

On conventional symmetric $L^4$ lattices at large $N$ there are
additional physical phases for $aL \leq 1/T_c$, corresponding to the
Polyakov loops acquiring non-zero expectation values one after the
other \cite{NarayananNeuberger}. The first of these, as one increase
$b$, is simply the usual deconfining transition and will occur at
$aL=1/T_c$. The subsequent transitions are expected to occur at
higher $b$ \cite{NarayananNeuberger} although there are unfortunately
no values in the literature for the corresponding critical values of
$b$. It is not clear why we have not observed any of these
transitions when we reduce $b$ from the twist-eating start,
particularly in the $N \leq 81$ case where the Eguchi-Kawai
correspondence should hold for all $b$. It may be that there is a
large metastability so that the twist-eating configuration tunnels to 
the bulk phase (for $N\leq 81$) or to the symmetry broken phase (for 
larger $N$) before it can tunnel to one of these physical `small 
volume' phases. Since the Eguchi-Kawai equivalence only holds for the 
planar limit, one cannot expect it to extend to the details of 
transitions and metastabilities which are typically driven by effects 
that are exponentially small in $N$, and so there may be some surprises 
here.

In this paper we have focused on two particularly interesting results
from our study of the twisted Eguchi-Kawai model: the symmetry
breaking that invalidates the Eguchi-Kawai planar correspondence over
a wide range of couplings, and the apparent inaccessibility of any
physical phase even at $N=144$. We have been able to make substantial
analytic progress in understanding these phenomena, using 
configurations that are a generalisation of the `fluxons' studied long 
ago by van Baal. Even if from a practical point of view our results 
undermine the utility of the most conventional TEK models, it would be 
theoretically interesting to pursue the numerical calculations to 
somewhat larger $N$, so as to obtain some insight into what happens at 
asymptotic $N$, and to develop further our analytic understanding of 
the model. This we hope to do elsewhere where we will also present our 
results for TEK models with anisotropic lattice spacings and for the 
partially reduced models that one would need to use for calculations of 
the planar mass spectrum.

\vspace*{0.30in}

%-------------------

\noindent {\bf Note added:} During the course of this work we became 
aware in \cite{OneLoopInstabitily} of an unpublished talk 
\cite{IshikawaOkawa} in which observations were made about centre 
symmetry breaking at intermediate couplings in TEK models.

\vspace*{0.30in}

%-------------------

\section*{Acknowledgements}
We are grateful to Barak Bringoltz for useful discussions and
suggestions. Our lattice calculations were carried out on PPARC and
EPSRC funded computers in Oxford Theoretical Physics. HV is supported
by FCT (Portugal) under the grant SFRH/BD/12923/2003.

%-------------------

\vfill\eject

\clearpage

%-------------------

%TABLES:

\begin{table}
\begin{center}
\begin{tabular}{|c|c|c|c|}\hline
$N$ & run & $n_{heats}$ & $n_{sweeps}$ \\\hline
64                   & run A & 200 & 6000 \\\hline
\multirow{2}{*}{81}  & run A & 200 & 6000 \\\cline{2-4}
                     & run B & 200 & 6000 \\\hline
\multirow{2}{*}{100} & run A & 200 & 6000 \\\cline{2-4}
                     & run B & 200 & 6000 \\\hline
121                  & run B & 100 & 2000 \\\hline
\multirow{2}{*}{144} & run A & 100 & 1000 \\\cline{2-4}
                     & run B & 100 & 1000 \\\hline
\end{tabular}
\caption{Parameters of the numerical simulations performed at each
value of the inverse 't Hooft coupling, $b$; $N$ is the size of the
gauge group; $n_{heats}$ is the number of thermalising sweeps;
$n_{sweeps}$ is the number of sweeps used for measurements.}
\label{table:SimulationDetails}
\end{center}
\end{table}

\begin{table}
\begin{center}
\begin{tabular}{|c|c|c|c|c|c|}\hline
$N$ & \multicolumn{5}{|r|}{$Z_N^{4(r)}~~~ \rightarrow ~~~~ Z_N^3 ~~~
\rightarrow ~~~~ Z_N^2 ~~~~ \rightarrow ~~~ Z_N^1 ~ \rightarrow ~ Z_N^0$}
\\\hline
\multirow{2}{*}{100} & run A & 0.305(5) & \multicolumn{3}{|c|}{---}
\\\cline{2-6}
& run B & 0.300(5) & \multicolumn{3}{|c|}{---} \\\hline
121 & run B & 0.275(5) & 0.350(5)& 0.44(1) & 0.445(5) \\\hline
\multirow{2}{*}{144} & run A & 0.26(2) & 0.325(5) & 0.405(15) & 0.44(2)
\\\cline{2-6}
& run B & 0.2575(25) & 0.3275(25) & 0.3975(25) & 0.425(5)
\\\hline
\end{tabular}
\caption{Critical values of the inverse 't Hooft coupling, $b$,
associated with the breaking of one (or more) $Z_N$ symmetries of the
TEK model. This table refers to simulations that started from
randomised configurations. $Z_N^k$ refers to a phase of the TEK model
with $k$ unbroken directions, while the arrows mean transitions
between those phases; a numerical value spanning multiple columns
corresponds to the critical coupling associated with the simultaneous
breaking of more than one $Z_N$ symmetry.
\label{table:TransitionsRand}}
\end{center}
\end{table}

\begin{table}
\begin{center}
\begin{tabular}{|c|c|c|c|c|c|c|}\hline
$N$ & \multicolumn{6}{|r|}{$Z_N^{4(r)}~ \leftarrow ~~ Z_N^3 ~~
\leftarrow ~~ Z_N^2 ~~ \leftarrow ~~ Z_N^1 ~~ \leftarrow ~ Z_N^0 ~
\leftarrow Z_N^{4(t)}$} \\\hline
\multirow{2}{*}{100} & run A & 0.27(1)  &
\multicolumn{4}{|c|}{0.350(5)} \\\cline{2-7}
& run B & 0.275(5) & \multicolumn{4}{|c|}{0.350(5)} \\\hline
121 & run B & 0.250(5) & \multicolumn{3}{|c|}{0.325(5)} & 0.360(5)
\\\hline
\multirow{2}{*}{144} & run A & 0.23(1)  & 0.27(1) &
\multicolumn{2}{|c|}{ ~~~~~ 0.325(5) ~~~~ } & 0.385(10) \\\cline{2-7}
& run B & 0.235(5)  & 0.275(5) & \multicolumn{2}{|c|}{ ~~~~~ 0.325(5)
~~~~ } & 0.370(5) \\\hline
\end{tabular}
\caption{Critical values of the inverse 't Hooft coupling, $b$,
associated with the breaking (or restoration) of one (or more) $Z_N$
symmetries of the TEK model. This table refers to simulations that
started from twist-eating configurations. $Z_N^k$ refers to a phase
of the TEK model with $k$  unbroken directions, while the arrows mean
transitions between those phases; a numerical value spanning multiple
columns corresponds to the critical coupling associated with the
simultaneous breaking (or restoration) of more than one $Z_N$
symmetry.\label{table:TransitionsCold}}
\end{center}
\end{table}

\vfill\eject

\clearpage

%-------------------

%FIGURES:

\begin{figure}[p]
\begin{center}
\leavevmode
\input{su64_plaq}
\end{center}
\vskip 0.015in \caption{Average value of the real part of the
plaquette, $\left\langle \mathrm{Re}~u_p\right\rangle$, in the SU(64)
TEK model versus the inverse bare 't Hooft coupling, $b$. The squares
($\square$) represent the $N\rightarrow\infty$ extrapolation of the
average plaquette in Wilson's lattice gauge theory. The solid line
(---) and the dashed line (- - -) represent the strong-coupling and
the 3-loop weak-coupling expansions of Wilson's lattice gauge theory,
respectively. \label{fig:SU64_plaq}}
\end{figure}

\clearpage

\begin{figure}[p]
\begin{center}
\leavevmode
\input{su100_link_cold}
\end{center}
\vskip 0.015in \caption{Average value of the real and imaginary
parts of traced link variables, $\frac{1}{N}\mathrm{Tr}~U_\mu$, in
the SU(100) TEK model versus the inverse bare 't Hooft coupling, $b$
(from a cold start simulation); the dot-dashed line (-- $\cdot$ --)
represents the average real plaquette. \label{fig:SU100_link_cold}}
\end{figure}

\clearpage

\begin{figure}[p]
\begin{center}
\leavevmode
\input{su100_link_hot}
\end{center}
\vskip 0.015in \caption{Average value of the real and imaginary
parts of traced link variables, $\frac{1}{N}\mathrm{Tr}~U_\mu$, in
the SU(100) TEK model versus the inverse bare 't Hooft coupling, $b$
(from a hot start simulation); the dot-dashed line (-- $\cdot$ --)
represents the average real plaquette. \label{fig:SU100_link_hot}}
\end{figure}

\begin{figure}[p]
\begin{center}
\leavevmode
\input{su144_plaq}
\end{center}
\vskip 0.015in \caption{Average value of the real part of the
plaquette, $\left\langle \mathrm{Re}~u_p\right\rangle$, in the
SU(144) TEK model versus the inverse bare 't Hooft coupling, $b$. The
squares ($\square$) represent the $N\rightarrow\infty$ extrapolation
of the average plaquette in Wilson's lattice gauge theory. The solid
line (---) and the dashed line (- - -) represent the strong-coupling
and the 3-loop weak-coupling expansions of Wilson's lattice gauge
theory, respectively. \label{fig:SU144_plaq}}
\end{figure}

\clearpage

\clearpage

\clearpage

\end{document}